\RequirePackage{amsmath}
\documentclass[12pt]{iopart}
\usepackage{cite}
\usepackage{graphicx}
\usepackage{wrapfig}
\usepackage{caption}
\usepackage{placeins}
\usepackage{mdframed}
\usepackage{wrapfig} 
\usepackage{siunitx}
\usepackage{mathtools}
\usepackage{aligned-overset}
\usepackage{amsfonts,amssymb}
\usepackage[x11names]{xcolor}
\usepackage{booktabs}
\usepackage[colorlinks=true,linkcolor=blue,citecolor=blue,filecolor=cyan,urlcolor=blue,breaklinks=true]{hyperref}
\usepackage[T1]{fontenc}
\usepackage{lscape}
\newcommand{\tcorr}{t_{\rm corr}}
\newcommand{\D}{\mathcal{D}_{\tau, q_0}}
\newcommand{\GCK}{G^{\rm CK}_\tau}

\newlength\InnerWidth
\begin{document}
\title[]{Model-free inference of memory in conformational dynamics of a multi-domain protein}

\author{Leonie Vollmar$^{1,2,*}$, Rick Bebon$^{3,*}$, Julia Schimpf$^{1,2}$, Bastian Flietel$^4$, Sirin Celiksoy$^4$, Carsten Sönnichsen$^4$, Alja\v{z} Godec$^3$ and Thorsten Hugel$^{1,5}$}

\address{$^1$ Institute of Physical Chemistry, University of Freiburg, 79104 Freiburg, Germany}
\address{$^2$ Spemann Graduate School of Biology and Medicine (SGBM), University of Freiburg}
\address{$^3$ Mathematical bioPhysics Group, Max Planck Institute for Multidisciplinary Sciences, 37077 Göttingen, Germany}
\address{$^4$ Department of Chemistry, Johannes Gutenberg-University of Mainz, 55128 Mainz, Germany}
\address{$^5$ Signalling Research Centers BIOSS and CIBSS, University of Freiburg, 79104 Freiburg, Germany}
\address{$^*$ contributed equally}
\ead{th@pc.uni-freiburg.de}
\date{\today}
\begin{abstract}
Single-molecule experiments provide insight into the motion (conformational dynamics) of individual protein molecules. Usually, a well-defined but coarse-grained intramolecular coordinate is measured and subsequently analysed with the help of Hidden Markov Models (HMMs) to deduce the kinetics of protein conformational changes. Such approaches rely on the assumption that the microscopic dynamics of the protein evolve according to a Markov-jump process on some network. However, the manifestation and extent of memory in the dynamics of the observable strongly depends on the chosen underlying Markov model, which is generally not known and therefore can lead to misinterpretations.   
Here, we combine extensive single-molecule plasmon ruler experiments on the heat shock protein Hsp90, computer simulations, and theory to infer and quantify memory in a model-free fashion.
Our analysis is based on the bare definition of non-Markovian behaviour and does not require any underlying model. In the case of Hsp90 probed by a plasmon ruler, the Markov assumption is found to be clearly and conclusively violated on 
timescales up to roughly $\SI{50}{\second}$, which corresponds roughly to $\sim$50\% of the inferred correlation time of the signal. The extent of memory is striking and reaches biologically relevant timescales.
This implies that  memory effects penetrate even the slowest observed motions. We provide clear and reproducible guidelines on how to test for the presence and duration of memory in experimental single-molecule data.\\

\end{abstract}
\submitto{\jpa}
\maketitle
\section{Introduction}
Dynamics of biological systems, including data obtained in single-molecule experiments, are often analysed by (Hidden) Markov Modelling \cite{mckinneyAnalysisSingleMoleculeFRET2006,gotzBlindBenchmarkAnalysis2022}. This method allows to infer mesoscopic states and Markovian transition rates between them from time series, which can be obtained from time-resolved distance measurements, e.g., by 
Förster resonance energy transfer (FRET) \cite{mckinneyAnalysisSingleMoleculeFRET2006, gotzBlindBenchmarkAnalysis2022}, plasmon rulers\cite{yeConformationalDynamicsSingle2018}, 
and optical \cite{pyo2019, stigler2011} or magnetic tweezers\cite{franzAllostericActivationVinculin2023}. Markov models have been very successful for analysing such data but their validity critically hinges on the assumption that the observed process is memoryless, meaning that  transitioning from one state to another is independent of previous states and that the dwell times within 
states are exponential random variables \cite{Hartich_PRX}. The latter assumption in particular also implies that the transition paths between meso-states are infinitely fast \cite{Hartich_PRX}.  
If either (or both) of the above assumption is violated, the interpretation in terms of Markov models may lead to inconsistent conclusions (see e.g.~\cite{Hartich_PRR}). 
Parametrically, memory effects in \emph{observed} dynamics can be incorporated by Hidden Markov models. However, the magnitude and duration of memory in the observed dynamics strongly depend on the choice of the underlying (microscopic) Markov model. 
Importantly,  
the magnitude and extent of memory in observed dynamics has not been consistently tested in any of these experimental studies. As a result, the sensibility and accuracy of Hidden Markov models of single-molecule observations \cite{mckinneyAnalysisSingleMoleculeFRET2006,gotzBlindBenchmarkAnalysis2022} has \emph{not} yet been scrutinised. 

Generally, memory is defined as the absence of Markov behaviour, and can emerge from `time lumping' or `state lumping' \cite{Hartich_PRX,Hartich_PRR,Grigolini,Haken, Haenggi_1, Haenggi_2, Blom2024}.
Time lumping directly results from biophysical measurements as the time resolution is not infinite and the signal is binned over small time windows. State lumping, on the other hand, refers to the states of the observed system, and also arises as a consequence of the measurement procedure: if an intramolecular distance of a biomolecule, for example, is observed over time by single-molecule FRET (smFRET) or plasmon ruler, the experimental information refers to only one spatial coordinate, which is the `projection' of many underlying conformational states. Memory can emerge when two prerequisites are met: first, when there are more degrees of freedom than those observed, signifying that an observed state is a projection of several coordinates onto e.g.~only one. Second, when \emph{dynamically distinct} microscopic states become projected onto the same observable state (for a specific \emph{minimal} example on how memory may emerge upon projection of `microscopic' Markov-jump dynamics see Box~1 as well as \cite{zhao_PRL}).
This holds true in the aforementioned example of the biomolecule as it consists of numerous atoms (i.e., it has numerous degrees of freedom) and exhibits (well-defined) long-lived conformational meso-states, but only the distance between two amino acids is observed. In this situation, memory is automatically induced unless all hidden degrees of freedom relax infinitely fast, such that the observable `feels their presence only on average'. 

One important question emerging from this consequential relation is whether such memory effects are biologically relevant. Therefore, it is crucial to quantify both their magnitude and duration. If they relax much faster than the observed biological processes, they can be safely neglected. Conversely, if they relax on a slower timescale, then they may directly affect biological processes as, e.g., protein folding, catalysis or signalling. In practice, this is hardly ever known a priori, and neither the exact microscopic dynamics nor their projection 
(i.e., the mapping between full and projected dynamics) 
are experimentally accessible. This in turn poses the challenging question \emph{how to detect and quantify memory without knowing the underlying microscopic dynamics}?

Non-exponential dwell time distributions in given meso-states in principle (i.e., if present) serve as an indicator for the presence of memory. However, an exponential dwell (i.e., exit time) distribution in turn is \emph{not} an indicator for Markovian dynamics, as memory can also emerge in the sequence of consecutively visited states (see example in \cite{Hartich_PRR}). 

A preliminary upper bound on the duration of memory can be inferred from the correlation (or `mixing') time $t_{\rm corr}$ of the signal. 

The (auto)correlation time may be defined via the (auto)correlation function $C(t)=\langle q_tq_0\rangle-\langle q_t\rangle\langle q_0\rangle$, which is determined as the average over a (large) ensemble of $N$ observed/projected trajectories or the time average over a sufficiently 
long trajectory, i.e.,
\begin{align}
C(t)&= \lim_{N\to\infty}\left(N^{-1}\sum_{i=1}^N q^i_t q^i_0-N^{-2}\sum_{i=1}^N q^i_t\sum_{j=1}^N q^j_0\right)\nonumber \\
&=\lim_{T\to \infty; T/t\gg 1}(T-t)^{-1}\int_0^{T-t}q_{\tau+t}q_\tau d\tau-\left(\lim_{T\to \infty}(T)^{-1}\int_0^{T}q_\tau d\tau\right)^2,
\label{corrf}
\end{align}
where we tacitly assume that the observed process $q_t$ is ergodic and that the \emph{microscopic} dynamics 
are stationary and time-homogeneous (i.e., that microscopic dynamics do \emph{not} explicitly depend on time).
The autocorrelation $C(t)$ of a time-series measures the correlation between the coordinate $q$ at $t=0$ and different time points $t$. In essence, it quantifies the `degree of similarity' between these lag times. 
As such, it captures the extend of `memory' of both the observable and hidden initial conditions.
The correlation time is defined as (see e.g.~\cite{lapollaToolboxQuantifyingMemory2021})
\begin{align}
\frac{1}{t_{\rm corr}}\equiv -\lim_{t\to\infty}\frac{1}{t}\ln C(t)
\label{corr-t}
\end{align}
and reflects the timescale on which the entire system (i.e., observable plus any hidden degrees of freedom) `forgets' the initial condition. Alternatively, one can instead define a normalised autocorrelation function $\mathcal{C}(t)\equiv C(t)/C(0)$.
Such an analysis is very helpful to obtain a first upper bound on the memory timescale, since (by definition) if memory exists the correlation time is the slowest timescale in the system.  However, it is important to note that the correlation time does not equal the memory time (see e.g.~\cite{Lapolla_2020}). In fact, it does not even imply the presence of memory. To see this, recall that a one-dimensional diffusion in a bi-stable or even in a single-well parabolic potential both have a non-vanishing correlation time  while they are perfectly Markovian. Moreover, conclusions drawn form so-called `normalised correlation times' \cite{Perico_1993,Kalmykov_2006}, $\hat{t}_{\rm corr}\equiv \int_0^\infty \mathcal{C}(\tau)d\tau$, 
may lead to misconceptions about the underlying dynamics when the latter span multiple timescales  \cite{Lapolla_2020}.
A correlation time analysis is also not suitable for extracting the temporal extent of memory  when the latter is present, because it is only an upper bound and therefore has no implications on the duration of memory. 

An elegant indicator of memory in low-dimensional reaction coordinates focuses on the fluctuations of transition-path times \cite{Berezhkovskii_2018} (see also generalisation for dynamics on general graphs in \cite{Hartich_PRX}). This analysis exploits the fact that transition-path times---the durations of successful transitions---for a one-dimensional Markov process cannot have a coefficient of variation exceeding one. Thus, values exceeding one are a reliable indicator of memory. However, a coefficient smaller than one does not necessarily imply Markov behaviour and this analysis also requires superb temporal resolution (much faster than the transition duration), which is often not feasible experimentally.

A more robust (concerning temporal resolution) and model-free (but still not ideal; see below) test for the presence and duration of memory is to test \emph{directly} for the violation of Markov behaviour~\cite{lapollaToolboxQuantifyingMemory2021}. This method offers significant flexibility and interpretability, free from the model-selection challenges \cite{hastie2009elements} inherent to model-based approaches like Hidden Markov models. The approach rests on one general consequence of the Markov property encoded in the so-called Chapman-Kolmogorov equation. That is, the (two-point) transition probability density $q_0\to q$ of any Markov process satisfies
\begin{align*}
G(q,t|q_0,0)\overset{{\rm Markov}}{=}&\int dq' G(q,t|q',\tau)G(q',\tau|q_0,0)\\
\overset{t-{\rm hom}}{=}&\int dq' G(q,t-\tau|q',0)G(q',\tau|q_0,0)
\end{align*}
where $t-$hom denotes time-homogeneous Markov dynamics. Note in particular that there is no dependence on $\tau$. 
Since the Chapman-Kolmogorov equation holds true for all Markov processes, one can turn this property 
into a test for Markovianity by determining the transition probability density $q_0\to q$ for the actual \emph{observed} (generally non-Markovian) dynamics,
$G(q,t|q_0,0)$, 

and compare it with the (fictitious) Chapman-Kolmogorov construction
\begin{equation}
    \GCK(q,t|q_0) \equiv \int dq' G(q,t-\tau|q',0) G(q', \tau|q_0,0) .
    \label{eq:GK_Construction}
\end{equation}
The Chapman-Kolmogorov construction $\GCK$ corresponds to a fictitious process in which we force all hidden degrees of freedom
to their stationary distribution at time~$\tau$, and thereby erase all memory (if present). 
The transition probability density can be determined from the recorded (experimental or simulated) 
time series $\{q_t\}_{0\leq t\leq T}$ (one-dimensional in our case) of total duration $T=\SI{6}{\hour}$ (e.g., see Fig.~\ref{fig:SimTraces+Corr}d-f or Fig.~\ref{fig:ExpTraces+Corr}a). 
Assuming that the time-series is stationary (which we verified to be the case here by confirming that waiting times in the states do \emph{not} explicitly depend on time), we can determine the 
two-point transition probability density as
\begin{align}
G(q,t|q_0,0)=\frac{\langle \delta(q_{t+\tau}-q)\delta(q_{\tau}-q_0)\rangle}{\langle \delta(q_{\tau}-q_0)\rangle},
\label{Greens}
\end{align}
where as before $\langle\cdot\rangle$ denotes an ensemble or time average over a long trajectory and $\delta(x)$ is Dirac's delta distribution. 
Due to our assumption that $q_t$ is ergodic, the system reaches a stationary equilibrium density $p_{\rm eq}(q)$ for ergodically long times, i.e. $\lim_{t\to\infty} G(q, t|q_0,0)= p_{\rm eq}(q)$.
Note that, given sufficient statistics, one can infer $G(q,t|q_0,0)$ and $\GCK(q,t|q_0)$ directly from the measured time series $q_t$, and that in particular, there is no need to specify any underlying 
model for the observed (projected) and microscopic dynamics. The comparison between $G$ and $\GCK$ may be done via the Kullback-Leibler divergence (also called relative entropy) \cite{kullbackInformationSufficiency1951}
\begin{equation}
    \D(t) \equiv \int dq G(q,t|q_0,0) \ln\left[\frac{G(q,t|q_0,0)}{\GCK(q,t|q_0)}\right]\ge 0.
    \label{klddiv}
\end{equation}
By construction, $\D(t)=0$ if and only if $G(q,t|q_0)=\GCK(q,t|q_0)$ for all $q$. That is, $\D(t)$ will by construction vanish for a Markov process.   
A non-zero value of $\D(t)$ for any $\tau$ and $t$ thus reflects genuine memory in the dynamics in the sense that as soon as $\D(t)\ne 0$ for some $\tau$ and $q_0$ the observed process $q_t$ is conclusively non-Markovian. Note, however, that there are non-Markovian processes which also satisfy the Chapman-Kolmogorov equation
\cite{lapollaManifestationsProjectionInducedMemory2019,Feller,Hanggi_CK} -- and thus would also result in a vanishing $\D$. Therefore, while $\D(t)\ne 0$ implies memory,  the converse $\D(t)=0$ does \emph{not} necessarily imply that the observed process is Markovian.

Notably, for long times $t\geq \tcorr$, and similarly for $\tau \geq \tcorr$ (and $t=\tau$ by construction), one expects that $\D(t)$ approaches zero. Analysing $\D(t)$ for different 
$\tau$ and $q_0$ thus allows us to quantify the magnitude and temporal extent of memory in a model-free manner (see below). The Kullback-Leibler divergence in Eq.~\eqref{klddiv} reflects how much the observable $q_t$ at a given time $t$ `remembers' the initial state of \emph{hidden} degrees of freedom and quantifies how memory attenuates in the course of time while microscopic trajectories gradually mix in configuration space.
Lastly, it is important to highlight that the information 
encoded in the temporal evolution of $\mathcal{C}(t)$ and $\D(t)$ is fundamentally different. 
This distinction becomes evident in Figs.~\ref{fig:SimTraces+Corr} and~\ref{fig:KL_sim}, 
which compare a Markovian with non-Markovian observations.
Even in the absence of memory, there exists a timescale on which the coordinate $q_t$ forgets the initial conditions. However, in the Markovian case, there is no memory and $\D(t)=0$ as required. Therefore, $\D(t)$ in contrast to $\mathcal{C}(t)$, indeed provides information about the correlation between the observable $q_t$ and the initial conditions of \emph{hidden} degrees of freedom, thereby offering genuine information about the extent and duration of memory.

\begin{figure}
    \centering
    \includegraphics[width=\textwidth]{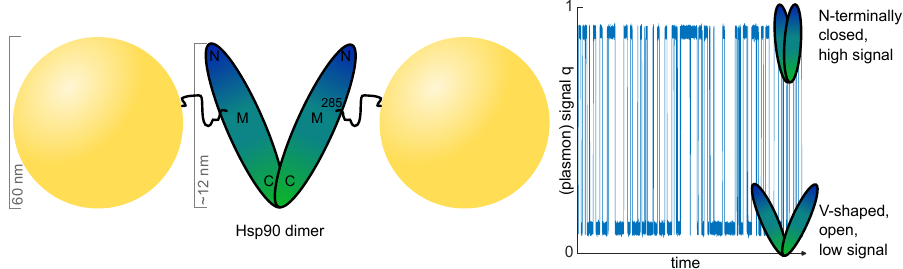}
    \caption{The heat shock protein Hsp90 (green-blue) is a homodimer composed of two identical proteins, which each have a N, M and C-domain. They are bound to each other at their C-termini, respectively. The protein exhibits N-terminal (top) opening and closing dynamics. This motion of a single dimer can be recorded by plasmon ruler spectroscopy in a dark field microscope where two gold nanoparticles are attached to the protein. As the coupling strength of the gold particles depends on their distance, a smaller distance results in stronger coupling and therefore larger scattering intensity.}
    \label{fig:Hsp90illu}
\end{figure}

Here, we investigate memory effects in single-molecule time series of the conformational dynamics of the protein Hsp90 probed by a plasmon ruler, monitoring the distance between a pair of plasmonic nanoparticles attached to the protein (see Fig.~\ref{fig:Hsp90illu}), as well as a set of hidden Markov models of this observable. Hsp90 is a molecular chaperone (a so-called `helper protein'), as such, it helps other proteins to fold properly and to achieve an active (i.e., biologically functional) conformation. As a heat shock protein, its production is modified by stress conditions.  Hsp90 is highly conserved and one of the most abundant proteins within the cytoplasm of cells~\cite{chenComparativeGenomicsEvolution2006,laiQuantitationIntracellularLocalization1984, moulickAffinitybasedProteomicsReveal2011}. Renowned for its extensive interaction network with numerous cochaperones (one may call them `helper's helpers') and protein clients (also called substrates), Hsp90 plays a pivotal role in the maintenance of essential cellular functions \cite{borkovichHsp82EssentialProtein1989, Schopf2017}. Structurally, Hsp90 exists as a homodimer, with each monomer composed of three domains, as shown in Fig.~\ref{fig:Hsp90illu}: the N-terminal domain (NTD) containing an ATP-binding site, the middle domain (MD) crucial for ATP-hydrolysis and linked to the NTD via a charged linker sequence, and the C-terminal domain (CTD), serving as the dimerisation interface with the highest affinity~\cite{pearlStructureFunctionMechanism2001,hoterHSP90FamilyStructure2018}. The Hsp90 dimer can undergo large conformational changes, transitioning from a V-shaped N-terminally open structure to a tightly closed state, as evidenced by a wealth of structural data~\cite{shiauStructuralAnalysisColi2006, aliCrystalStructureHsp90nucleotidep232006}. Several investigations have measured time series of N-terminal opening and closing with single-molecule 
FRET~\cite{micklerLargeConformationalChanges2009, schmidSingleMoleculeAnalysisDwell2016, schmidControllingProteinFunction2020, vollmarCochaperonesConveyEnergy2024}. 
These studies are limited by photobleaching, resulting in insufficient statistics to asses the Markov assumption. 
However, plasmon ruler spectroscopy allows for a much longer observation of Hsp90’s conformational changes, monitoring its opening and closing dynamics in the range of hours~\cite{yeConformationalDynamicsSingle2018}. In the following, we analyse some of the plasmon ruler data from Ref.~\cite{yeConformationalDynamicsSingle2018}. The plasmon data is available for review at: https://nas.physchem.uni-freiburg.de:5679/sharing/1qUUz6f6s and will be provided on Zenodo upon acceptance of the manuscript.

Signatures of memory in the time-series of Hsp90 were already observed in Ref.~\cite{yeConformationalDynamicsSingle2018} in terms of non-exponential dwell times in the two respective conformational meso-states. Non-Markovian behaviour is indeed anticipated to manifest since many protein conformations lead to the same observable plasmon signal $q_t$, here a proxy for the distance between plasmonic particles, while microscopic trajectories evolving from an ensemble of conformations with the same distance between plasmonic particles will be dynamically distinct. The magnitude and timescale on which memory emerges remain unknown.   

\section{Simulated single-molecule time series}
\subsection{Data simulation}
\begin{figure}
    \centering
    \includegraphics{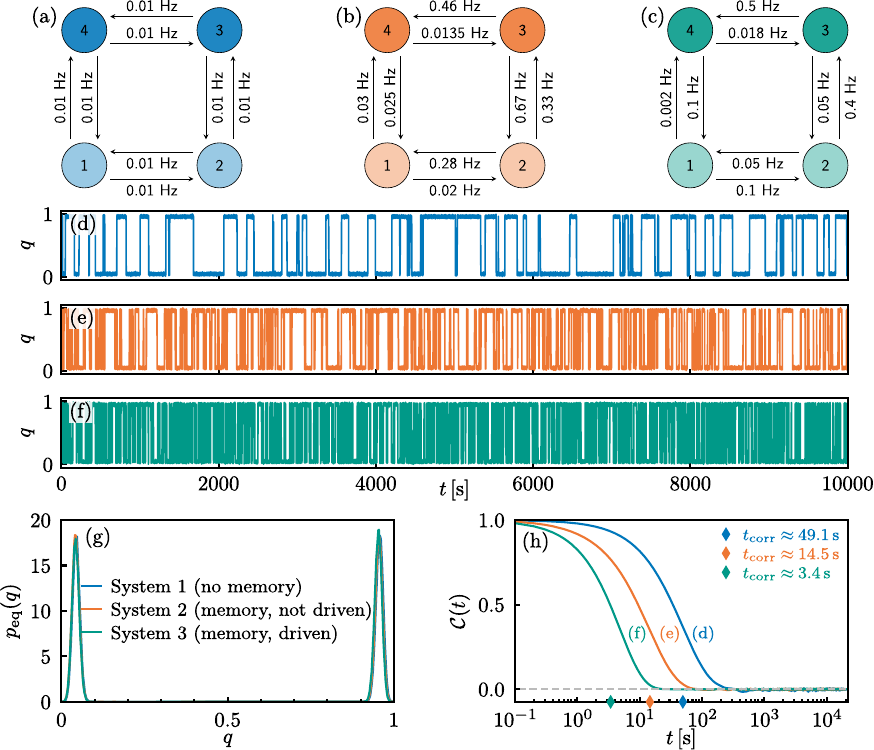}
    \caption{Examples of simulated time-series for three different 4-state models projected onto 2 observable states.
Each time-series $\{q_t\}_{0\leq t\leq T}$ of total duration $T=\SI{6}{\hour}$ (only \SI{10000}{\second} depicted)
is shown for a 4-state Markov model with (a+d)~uniform transitions, 
(b+e)~with Hsp90-inspired transition dynamics that obey detailed balance, 
(c+f)~with Hsp90-inspired transition dynamics that is driven out of equilibrium.
(g)~Probability density functions $p_{\rm eq}(q)$ of one example time-series for each of the introduced model systems (almost lying on top of each other).
The two peaks correspond to the closed (higher $q$) and open (lower $q$) 
conformation.
(h) Averaged autocorrelation function $\mathcal{C}(t)$ of simulated 
time-series of the corresponding models. 
The timescale $\tcorr$ is identified from the long-time behaviour of $\mathcal{C}(t)$ as detailed in the main text.}
\label{fig:SimTraces+Corr}
\end{figure}

In order to examine the consistency and limitations of this approach, which requires lots of data, 
we start with simulated time-series of model systems both with and without memory, and in or driven out of equilibrium. 
All the three models consist of four states each (Fig.~\ref{fig:SimTraces+Corr}a-c) and are
described by Markovian jump dynamics on the microscopic level.

Moreover, we consider dynamics on a discrete time, `probed' with a frequency of $\SI{10}{\Hz}$, since in the practical
context of biophysical measurements (see Sec.~\ref{sec:exp_analysis}) the recordings are inherently discrete
(which here in the limit of long measurements is expected to essentially become equivalent to continuous-time dynamics, e.g.~Gillespie simulations \cite{Celebration1, Celebration2}).
The simulation of time-series
for the full dynamics was done with MASH-FRET \cite{bornerSimulationsCamerabasedSinglemolecule2018}, 
which includes (Poissonian) photon shot noise, similar to the noise occurring in single-molecule FRET time series, yielding a signal that is \emph{not} dichotomous anymore.

The respective model parameters (in particular signal intensities) were chosen to mimic the distinct protein conformations observable in the experiments.
Corresponding kinetic transitions rates between states were either chosen to be identical
(Fig.~\ref{fig:SimTraces+Corr}a),
or inspired by those recently obtained from single-molecule FRET data of Hsp90’s conformational dynamics (Fig.~\ref{fig:SimTraces+Corr}b-c) \cite{vollmarCochaperonesConveyEnergy2024}. 
In the latter case, the first system (see Fig.~\ref{fig:SimTraces+Corr}b) models Hsp90 dynamics in equilibrium (i.e., transition rates obey detailed balance)
whereas the second system (see Fig.~\ref{fig:SimTraces+Corr}c) is out of equilibrium 
with an affinity of $10~k_\mathrm{B}T$ \cite{Seifert}.

To allow for a better comparison
with experimental data in later sections, all simulated 4-state models
were divided into two low intensity (centred at $q=0.1$; open) and two high intensity (centred at $q=0.9$; closed) states
which are consecutively lumped together to create hidden states, giving rise to only two observable meso-states.
The resulting `projected' dynamics $q_t$ (Fig.~\ref{fig:SimTraces+Corr}d-f) 
thus now mimic the effective 2-state dynamics observed in the experimental setup 
(compare Fig.~\ref{fig:ExpTraces+Corr}a).
Recall that the noise introduced by MASH-FRET blurs the discrete-state trajectory, ultimately giving meso-states a finite extension as seen in 
the stationary histograms $p_{\rm eq}(q)$ in Fig.~\ref{fig:SimTraces+Corr}g.

The following memory analysis will be carried out for data sets comprising 
100 independent time-series for each system with a duration of $T=\SI{6}{\hour}$ and time resolution of $\SI{10}{\Hz}$ each---identical parameters as in the experimental case.

The simulation and data analysis were performed by independent researchers. The person analysing the data thus had no prior knowledge of the underlying system.

\subsection{Analysis of simulated single-molecule time series}

We start our analysis by first determining the normalised autocorrelation function $\mathcal{C}(t)$ (see Eq.~\eqref{corrf}) for each individual time-series
of the three introduced models, and consecutively take the corresponding sample-average
over the respective 100 independent measurements (see Fig.~\ref{fig:SimTraces+Corr}h). 
Furthermore, the correlation times $t_{\rm corr}$  are determined by 
fitting the long-time behaviour of $\mathcal{C}(t)$ to an exponential $\mathcal{C}(t) \simeq \exp(-t/t_{\rm corr})$.
The inferred values are $\tcorr\approx \SI{49.1}{\second}$, $\tcorr\approx\SI{14.5}{\second}$, and $\tcorr\approx\SI{3.4}{\second}$, respectively, as indicated by the markers in Fig.~\ref{fig:SimTraces+Corr}h
(further details are given in Tab.~\ref{tab:sim}).

\begin{figure}[h!tb]
    \centering
    \includegraphics{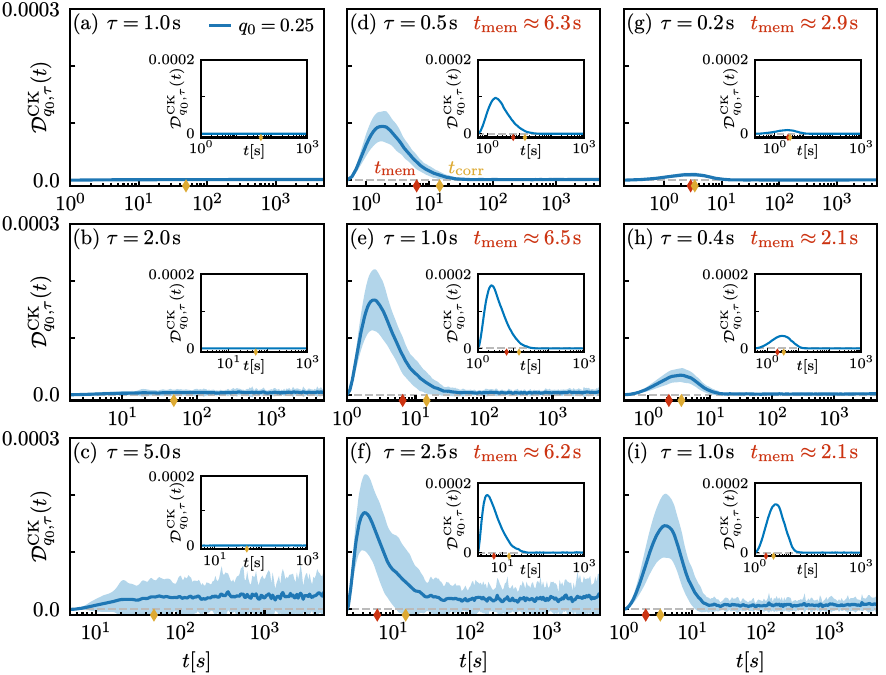}
    \caption{   
Analysis of memory in projected dynamics for three different simulated model systems
with 100 time-series of $\SI{6}{\hour}$ duration each.
The Kullback-Leibler divergence $\D(t)$ in Eq.~\eqref{klddiv} between the
transition probability density of the observed dynamics
$G(q,t|q_0)$ and its Chapman-Kolmogorov construction $\GCK(q,t|q_0)$
as a function of time $t$ for underlying dynamics of 
(a-c) a 4-state Markov model with uniform transitions, 
(d-f) a 4-state Markov model with Hsp90-inspired transition dynamics that obey detailed balance, and
(g-i) a 4-state Markov model with Hsp90-inspired transition dynamics that is driven out of equilibrium, respectively.
Initial condition was set here to $q_0=0.25$, corresponding to the open 
conformation of Hsp90 (see appendix for $q_0=0.75$).
Memory in the Chapman-Kolmogorov construct is reset to zero 
at different times $\tau$ and the memory time-scale $t_{\rm mem}$ (red marker) is 
obtained by fitting the long-time behaviour according to $\D(t)\simeq \exp(-t/t_{\rm mem})$. 
Shaded areas depict standard deviations $\sigma_\mathcal{D}(t)$, obtained
by averaging over $100$ independent simulated trajectories
and insets depict the same analysis for only one trajectory that is 100 times as long.}
\label{fig:KL_sim}
\end{figure}

Next we quantified the memory via the Chapman-Kolmogorov construction~\eqref{eq:GK_Construction} 
(see Fig.~\ref{fig:KL_sim}). 
The transition probability density $G(q,t|q_0,0)$ for the recorded time-series was determined according to Eq.~\eqref{Greens}
via a histogram analysis. Since the probability density functions $p_{\rm eq}(q)$ of all considered models (also for the experimental data; see Sec.~\ref{sec:exp_analysis}) are bimodal, 
indicating two well-defined and well-separated meso-states (see Fig.~\ref{fig:Hsp90illu}), we start by lumping the state-space into two equal bins of width $l_{\rm bin}=0.5$, centred at $0.25$ and $0.75$ respectively. 
As a result, $q$ is now discrete-valued, $q=\{0,1\}$, and refers to a pair of bins at $0.25$ and $0.75$ with a  width $l_{\rm bin}=0.5$. Each bin encompasses one of the two wells,
such that $q_0=0.25$ corresponds to the open state and $q_0=0.75$ to the closed state throughout, respectively.
Note that the precise position of $q$ is thus immaterial.
Consequently, the reference Chapman-Kolmogorov construction and in turn also the Kullback-Leibler divergence $\D(t)$ are straightforwardly obtained via Eqs.~\eqref{eq:GK_Construction} and \eqref{klddiv} by replacing integrals by sums over the two lumps.
For the following analysis, we choose $q_0=0.25$ and for completeness show the corresponding results for $q_0=0.75$ in the Appendix 
since they are almost identical.

\begin{figure}[h]
    \centering
    \includegraphics{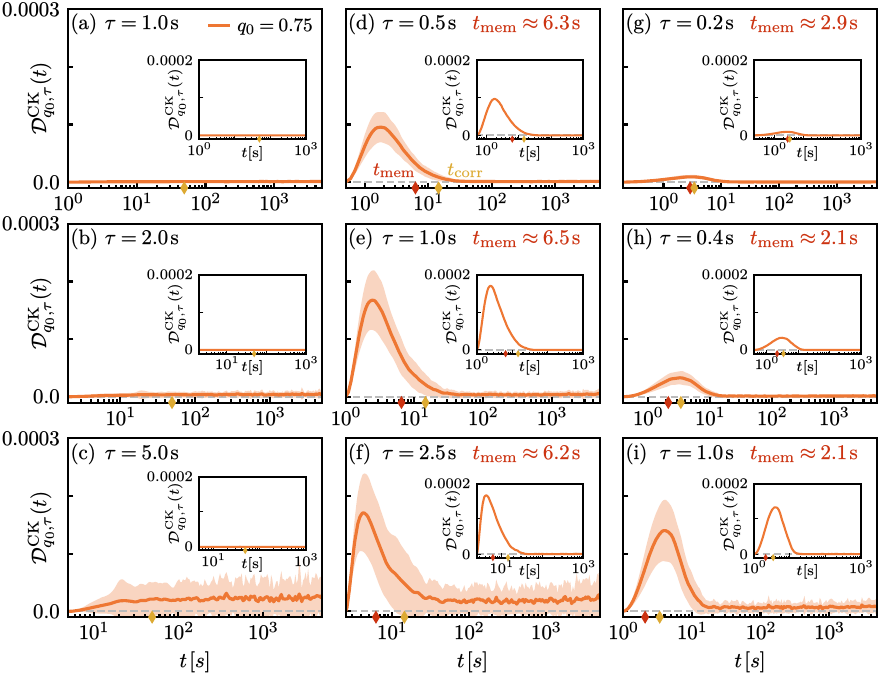}
    \caption{
Analysis of memory in projected dynamics for three different simulated model systems
with 100 time-series of $\SI{6}{\hour}$ duration each.
The Kullback-Leibler divergence $\D(t)$ in Eq.~\eqref{klddiv} between the
transition probability density of the observed dynamics
$G(q,t|q_0)$ and its Chapman-Kolmogorov construction $\GCK(q,t|q_0)$
as a function of time $t$ for underlying dynamics of 
(a-c) a 4-state Markov model with uniform transitions, 
(d-f) a 4-state Markov model with Hsp90-inspired transition dynamics that obey detailed balance, and
(g-i) a 4-state Markov model with Hsp90-inspired transition dynamics that is driven out of equilibrium, respectively.
Initial condition was set here to $q_0=0.75$, corresponding to the closed 
conformation of Hsp90 (see main text for $q_0=0.25$).
Memory in the Chapman-Kolmogorov construct is reset to zero 
at different times $\tau$ and the memory time-scale $t_{\rm mem}$ (red marker) is 
obtained by fitting the long-time behaviour according to $\D(t)\simeq \exp(-t/t_{\rm mem})$. 
Shaded areas depict standard deviations $\sigma_\mathcal{D}(t)$, obtained
by averaging over $100$ independent simulated trajectories
and insets depict the same analysis for only one trajectory that is 100 times as long.}
\label{fig:KL_sim_appendix}
\end{figure}

A priori, the
consistency of our approach requires that
\emph{no} detection of memory is expected for the first system (Fig.~\ref{fig:SimTraces+Corr}a), since in this case 
all possible microscopic paths are equivalent (i.e., all transition rates are identical). 
Indeed, as anticipated, 
the Kullback-Leibler divergence $\D(t)$~\eqref{klddiv} was found to be zero (within statistical errors) for all times $t$ and shown values of $\tau$
(see Fig.~\ref{fig:KL_sim} first column). 
Only for increasingly large values of $\tau$, a noisy result was obtained due to undersampling of the functions $G$ entering the Chapman-Kolmogorov construction~\eqref{eq:GK_Construction} necessarily yielding a positive $\D$.
Note that the uncertainty emerging for large $\tau$ is due to undersampling, i.e. can be overcome by longer traces (see inset Figs.~\ref{fig:KL_sim} and~\ref{fig:KL_sim_appendix}).
We remark that while no memory was (correctly) identified, the system displayed a correlation time of $t_{\rm corr}\approx \SI{49.1}{\second}$, underscoring that the correlation time $\tcorr$ (yellow marker) is a bad proxy for inferring memory. It does provide an upper bound on the duration of memory if the latter is present, but in turn has no implication for the presence of memory.  

For the two remaining models
with rates inspired by single-molecule FRET data of Hsp90 dynamics in equilibrium 
(Fig.~\ref{fig:SimTraces+Corr}b) and driven out of equilibrium
with an affinity of $10$ $k_{\rm B}T$ (Fig.~\ref{fig:SimTraces+Corr}c),
an analogous analysis confirms the presence of memory.
In both instances, the system displays memory identified by a non-vanishing Kullback-Leibler divergence over approximately one decade in time (Fig.~\ref{fig:KL_sim} 2nd and 3rd column). For the system in equilibrium, memory was best detectable for a $\tau=\SI{1}{\second}$. 
As before, large values of $\tau$ gave rise to undersampling and in turn larger fluctuations (compare shaded regions), and the results become noisier and less reliable. 
When using longer time traces, more data points are available and this problem does not emerge (see
insets in Fig.~\ref{fig:ExpTraces+Corr} for single trajectories that are longer by a factor of 100). 
For practical applications, this means that longer measurements or experiments with a higher time resolution result in a more stable and reliable analysis. In particular, an analysis of long traces performs better than the same amount of data split into several short traces (all longer that the correlation time). This is because for a given lag time $t$ in Eq.~\eqref{Greens} one has a total interval $T-t$ available for averaging per trajectory of duration $T$, and thus a total interval $p\times (T-t)$ for $p$ trajectories of length $T$ and a total interval $p\times T-t$ for a single trajectory of length $p\times T$. If the correlation time is $t_{\rm corr}$, one thus has roughly $\approx p\times (T-t)/t_{\rm corr}$ statistically independent realisations with $p$ trajectories compared to $\approx (p\times T-t)/t_{\rm corr}$ statistically independent realisations with a single long trajectory.

In line with the correlation time, one may analogously introduce a memory time-scale $t_{\rm mem}$ as 
\begin{align}
 \frac{1}{t_{\rm mem}}\equiv -\lim_{t\to\infty}\frac{1}{t}\ln \D(t),
\end{align}
which is inferred by fitting the longest time-scale to an exponential
$\D(t)\simeq \exp(-t/t_{\rm mem})$.
Consistently, we always find $t_{\rm mem} \leq \tcorr$ where, for the considered parameters of $\tau$, 
the equilibrium system was found to take values in the range of $t_{\rm mem}\approx\SI{6.2}{\second}$
to $t_{\rm mem}\approx\SI{6.5}{\second}$
and the driven system comparably $t_{\rm mem}\approx\SI{2.1}{\second}$ 
to $t_{\rm mem}\approx\SI{2.9}{\second}$ (see Fig.~\ref{fig:KL_sim} and Tab.~\ref{tab:sim}).

\section{Experimental single-molecule time series}
\label{sec:exp_analysis}
\subsection{Experimental data acquisition}
The experimental data analysed here has been recorded by plasmon ruler spectroscopy. For these measurements, two gold nanoparticles (diameter about 60~nm) were attached to yeast Hsp90 via flexible  PEG (polyethylene glycole) polymer linkers, one at each monomer. For the attachment of the PEG, Hsp90 was mutated for a cysteine at position 285 in the M-domain. These constructs were  immobilised on a glass substrate and observed in a dark field microscope. Due to plasmon coupling between the two gold nanospheres, their plasmon resonance is shifted to higher wavelengths, and the scattering intensity increases when the interparticle distance is decreased~\cite{Reinhard2005}. The scattering intensity is recorded and allows to monitor one spatial coordinate of Hsp90’s conformational changes. The open Hsp90 dimer shows a low plasmon scattering intensity in arbitrary units, while the closed dimer gives a high signal (see Fig.~\ref{fig:Hsp90illu}). 
We analysed four time series of duration $T=\SI{6}{\hour}$ each with a recording rate of 10 Hz from previously published data~\cite{yeConformationalDynamicsSingle2018}. 
All traces were from Hsp90 in equilibrium, i.e. in the absence of nucleotides.
One exemplary trace is shown in Fig.~\ref{fig:ExpTraces+Corr}a.
The effect of the coupling of the gold nanoparticles to the here investigated Hsp90's dynamics could already be excluded in the original publication~\cite{yeConformationalDynamicsSingle2018}, with the diffusion time of $\SI{0.1}{\micro \second}$ for 
$\SI{60}{\nm}$ gold nanoparticles being around six orders of magnitude faster than the observed Hsp90 dynamics here. 

\subsection{Reliability analysis via bootstrapping}
To gauge the reliability and robustness of the results with respect to (an unavoidably) limited sampling, we employ a simple bootstrapping routine. In addition to performing the analysis of the relative entropy in Eq.~\eqref{klddiv} inferred from the complete dataset, we repeated the analysis on bootstrapped sets. Precisely, we determined the relative entropy from  $M=100$ independent bootstrap samples obtained by randomly neglecting $20\%$ of the data. From the respective relative entropies of the bootstrapped samples $\D^{(i)}(t)$, we determine the standard deviation $\sigma_\mathcal{D}$ of $\D(t)$ as 
\begin{equation}
\sigma^2_\mathcal{D}(t) \equiv M^{-1} \sum_{i=1}^M \D^{(i)}(t)^2 - \left(M^{-1} \sum_{i=1}^M \D^{(i)}(t)\right )^2.
\end{equation}

\subsection{Analysis of experimental single-molecule time series}
\begin{figure}[h!tb]
    \centering
    \includegraphics{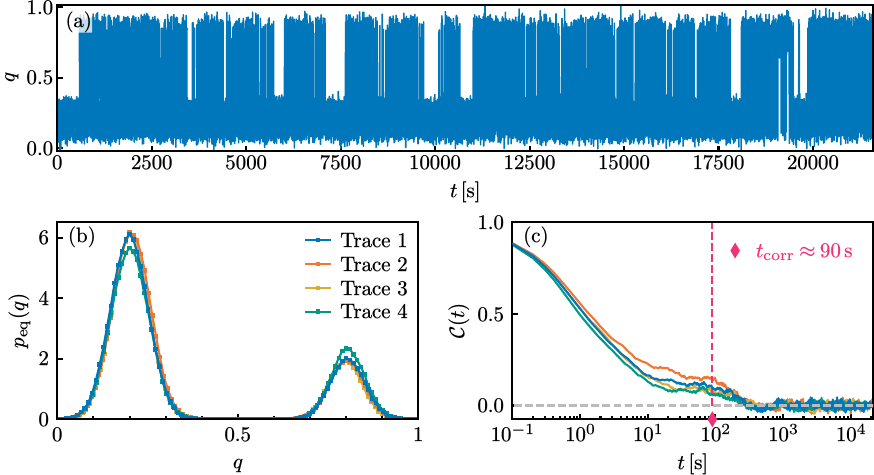}
    \caption{
Overview of experimental single-molecule time-series for the Hsp90 protein.
(a) Recorded example time-series $\{q_t\}_{0\leq t\leq T}$ of duration $T=\SI{6}{\hour}$ via 
plasmon ruler spectroscopy \cite{yeConformationalDynamicsSingle2018}.
(b) Probability density function $p_{\rm eq}(q)$ of four recorded Hsp90 time-series $q_T$.
The two visible peaks correspond to the closed (higher population) and open (lower population) 
conformation of the Hsp90 protein, respectively.
(c) Autocorrelation function $\mathcal{C}(t)$ of the recorded experimental time-series. The correlation
time $\tcorr$~(see Eq.~\eqref{corr-t}) is identified via an exponential fit of the long-time behaviour
of the (averaged) autocorrelation $\mathcal{C}(t)$.}
    \label{fig:ExpTraces+Corr}
\end{figure}

Fig.~\ref{fig:ExpTraces+Corr}a and b show a measured trajectory $q_t$ and the equilibrium probability density functions $p_{\rm eq}(q)$ of the four analysed trajectories,
respectively. As in the simulations, the two peaks correspond to the closed (high $q$) and open (low $q$) conformation of the Hsp90 protein. 
Note that the plasmon signal (here $q$) was normalised between zero and one and scaled such 
that the peak positions inferred from individual traces match.
The normalised autocorrelation functions $\mathcal{C}(t)$ of the four traces shown in Fig.~\ref{fig:ExpTraces+Corr}c are found to nicely superimpose and display an (averaged)
correlation time of $\tcorr\approx\SI{90}{\second}$ (see Tab.~\ref{tab:exp}), as above, 
obtained via an exponential fit of the long-time behaviour.

\begin{figure}[h!tb]
    \centering
    \includegraphics{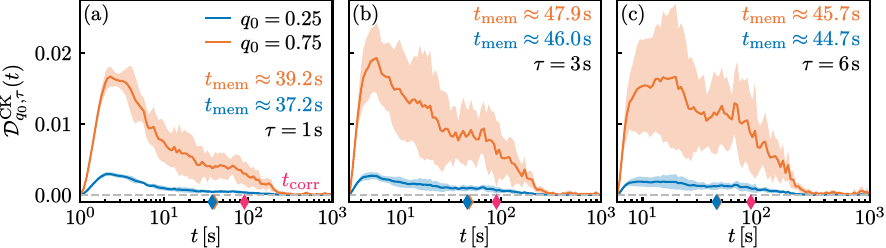}
    \caption{
Analysis of memory in experimental Hsp90 time-series.
Kullback-Leibler divergence $\D(t)$ in Eq.~\eqref{klddiv} 
between the
transition probability density of the observed dynamics
$G(q,t|q_0)$ and its Chapman-Kolmogorov construction
$\GCK(q,t|q_0)$
as a function of time $t$.
Two different initial conditions $q_0$ of the Hsp90 protein are considered,
corresponding to starting in the left (blue) and right (orange)
peak in the probability density $p_{\rm eq}(q)$, respectively.
Memory in the Chapman-Kolmogorov construction is reset to zero at times 
(a) $\tau=\SI{1}{\second}$, (b) $\tau=\SI{3}{\second}$, and $\tau=\SI{6}{\second}$, respectively.
Error bars depict the standard deviation $\sigma_\mathcal{D}(t)$, obtained
by systematically neglecting $20\%$ of the recorded data
for $M=100$ independent bootstrapped repetitions and subsequent averaging.
Memory time-scales are identified via an exponential fit of the long-time behaviour.
Note that $t_{\rm mem}$ (orange and blue markers) is always smaller than the correlation time $\tcorr$ (pink marker).}
    \label{fig:KL_exp}
\end{figure}

The Kullback-Leibler divergence for different `memory reset times' $\tau$ are shown in Fig.~\ref{fig:KL_exp} for different initial conditions $q_0$, respectively. Note that the memory profile, in particular the magnitude, depends on the initial condition of the observable. That is, when starting in the open state ($q_0=0.25$, blue curve in Fig.~\ref{fig:KL_exp}), the maximum of $\D(t)$ is smaller than when starting from the closed state ($q_0=0.75$, orange curve in Fig.~\ref{fig:KL_exp}). From a biophysical perspective, this is not too surprising, given the structural complexity of Hsp90 and the fact that the open state is much more flexible than the closed state \cite{hellenkampMultidomainStructureCorrelated2017}. In any case, the finding seems to indicate that a reduced model for $q_t$ in terms of memory-kernels must \emph{not} be of convolution type and should depend explicitly on $q_0$.

Signatures of memory emerged already at very short times (on the scale of the temporal resolution of the experiment). The maximum was reached after just a few seconds (Fig.~\ref{fig:KL_exp}). Then, the memory decreased slowly and was completely lost only after around $\SI{50}{\second}$. The memory timescale was again obtained from an exponential fit to the long-term behaviour
and was here $t_{\rm mem}\approx\SI{37.2}{\second}$ to $t_{\rm mem}\approx\SI{47.9}{\second}$, respectively, 
for different $\tau$ and $q_0$
(see Fig.~\ref{fig:KL_exp} and Tab.~\ref{tab:exp}). Thus, memory was present for over two decades and most pronounced for the `well' corresponding to the closed conformation of Hsp90 (orange in Fig.~\ref{fig:KL_exp}). This could be due to Hsp90 being more flexible in its open state \cite{hellenkampMultidomainStructureCorrelated2017} enabling a faster relaxation of hidden degrees of freedom. We stress that the actual memory profile in Fig.~\ref{fig:KL_exp} is (qualitatively and quantitatively) very different from the one predicted by the Hsp90-inspired Hidden Markov models in Fig.~\ref{fig:KL_sim}. In particular, the actual extent of memory is an order of magnitude longer than predicted by  Hsp90-inspired Hidden Markov models (e.g., $t_{\rm mem}\approx \SI{40}{\second}$ (actual) versus $t_{\rm mem}\approx \SI{6}{\second}$ (HMM) at $\tau=\SI{1}{\second}$)  and also displays (at least) two distinct time-scales (note the shoulder in $\D(t)$ shown in Fig.~\ref{fig:KL_exp}). 
Moreover, the Hidden Markov model 
falls short in accurately describing the correlation time ($\tcorr\approx\SI{90}{\second}$ (actual) versus $\tcorr\approx\SI{14.5}{\second}$ (HMM)), and fails to capture the observed clear 
distinction between the open (orange) and closed (blue) conformation.
The choice of the Hidden Markov model thus \emph{directly} affects the memory profile. A model that fails to capture the memory profile of the observed dynamics --- an intrinsic signature of the \emph{actual} coupling of the observable to hidden degrees of freedom --- does \emph{not} provide an adequate representation of the underlying microscopic dynamics.

Our analysis thus suggests that the transition between opening and closing occurs on timescales influenced by memory, and therefore is affected by the initial conditions of the hidden degrees of freedom.

\section{Discussion and Conclusions}
The conformational dynamics of proteins are often measured between two sites in a protein by single-molecule methods. The resulting time-series are then typically analysed by Hidden-Markov based methods, without properly assessing the representation of the coupling of the observable with hidden degrees of freedom that is parametrically encoded in the underlying \emph{full} Markov model.
Here, we analysed long single-molecule time-series of the conformational dynamics of the Hsp90 protein probed by a plasmon ruler. In particular, we carried out a model-free analysis of the non-Markovian behaviour and conclusively confirmed and quantified the memory in the observed dynamics. This was presumably the first time that the conformational dynamics of a protein were quantified in a model-free manner and in such detail in an experimental setting (model-based approaches have been used before \cite{Kou,Min,Hagen,Dalton_Netz_PNAS_2023,Glatzel_Schilling_EPL_2021}). Moreover, our analysis allowed us to quantify the lifetime of the memory in the time-series of Hsp90. Strikingly, the memory was confirmed to be present for up to $\SI{50}{\second}$. Recall that the effect of the coupling of the gold nanoparticles to Hsp90's dynamics on the investigated time scales could already be excluded in the original publication~\cite{yeConformationalDynamicsSingle2018}, such that we can attribute the memory to Hsp90 dynamics alone. The duration of memory is significant for many biological processes, therefore it is conceivable (if not very likely) that memory has biological relevance and implications. This means, for example, that interactions with client proteins are affected by Hsp90's conformational dynamics and vice versa. Thus, Hsp90’s interactions have long lasting effects and are influenced by its dynamics and interactions for up to $\SI{50}{\second}$. To set the timescale into perspective, human RNA is translated to a protein at a rate of approximately 5 amino acids per second~\cite{olofssonStructureBiosynthesisApolipoprotein1987}. For a protein consisting of 400 amino acids, translation would take about $\SI{80}{\second}$. If Hsp90 --- fulfilling its chaperone function --- assists in folding this protein into its mature form, the starting conformation of Hsp90 would directly affect how Hsp90 assists in the folding process. 
Furthermore, our analysis is also suitable for inferring memory in non-equilibrium (driven) systems as we have exemplified in Fig.~\ref{fig:KL_sim}. We anticipate that this will be very important in the future, as compared to Hsp90's ATPase rate (1 ATP per min~\cite{panaretouATPBindingHydrolysis1998,girstmairHsp90IsoformsCerevisiae2019}) memory of $\SI{50}{\second}$ is still long with likely significant impact on Hsp90's non-equilibrium dynamics.

Altogether, our study shows that for the analysis of single-molecule time-series, e.g., of protein dynamics, it is crucial to initially investigate whether memory is present. Only in the absence of memory on relevant timescales, Markov modelling is justified, because it relies on the assumption of a memoryless process. Additionally, (hidden) Markov modelling relies on assumptions about models and --- in the context of the model selection process and all its associated challenges \cite{hastie2009elements, Dalton_Netz_PNAS_2023,Meyer_Schilling_EPL_2019} --- one typically selects the one with the highest likelihood, potentially introducing bias and explicitly setting the memory profile. A Hidden Markov model that fails to correctly reproduce the memory profile (i.e., its magnitude and especially duration) \emph{cannot} be considered as an adequate representation of the underlying dynamics, in particular the coupling of the observable to hidden degrees of freedom, and thus cannot provide reliable insight.   
Model-free methods as described here, on the other hand, can overcome this bias given sufficient statistics.
Transition-path based approaches \cite{Satija_Makarov_PNAS_2020}  could then be used. Our data and analysis indicate a pressing need for such model-free methods that have to be further developed and optimised to analyse single-molecule time-series. On the positive side, our work demonstrates that such an analysis, despite requiring substantial statistics, is already within experimental reach.

\subsection*{Acknowledgements}

This work was supported by the Studienstiftung des Deutschen
Volkes (to RB), the Deutsche Forschungsgemeinschaft (DFG) under Germany's Excellence Strategy (CIBSS EXC-2189 Project ID 390939984, to TH), SFB1381 programme (Project ID 403222702, to TH) and European Research Council (ERC) under the European Union’s Horizon Europe research and
innovation programme (grant agreement No 101086182
to AG) as well as the German Research Foundation (DFG)
through the Heisenberg Program (grant GO 2762/4-1
to AG).

\subsection*{Data availability statement}

All data is available at the following link for review: https://nas.physchem.uni-freiburg.de:5679/sharing/1qUUz6f6s . Upon acceptance of the manuscript the data will be available on Zenodo.

\newpage

\section*{Appendix: Fit parameters for correlation time $t_{\rm corr}$ and memory time $t_{\rm mem}$}
\begin{table}[h]
\caption{Correlation timescale $\tcorr$ and memory timescale $t_{\rm mem}$ for the three different simulated model systems (here $q_0=0.25$ and different $\tau$).
Values are obtained by fitting the long-time behaviour of $\mathcal{C}(t)$ and $\D(t)$, respectively, 
to the exponential $\simeq \exp(-t/t_i)$ where $i\in\{{\rm corr}, {\rm mem}\}$.}
\label{tab:sim}
\begin{tabular}{ccccc}
\toprule
\multicolumn{1}{c}{System} & \multicolumn{1}{c}{$t_{\rm corr}$\,[s]} & \multicolumn{1}{c}{$t_{\rm mem}$\, [s]} & \multicolumn{1}{c}{$t_{\rm mem}$\, [s]} & \multicolumn{1}{c}{$t_{\rm mem}$\, [s]} \\
& & $(\tau=\SI{1.0}{\second})$ & $(\tau=\SI{2.0}{\second})$ & $(\tau=\SI{5.0}{\second})$\\
\cmidrule[0.4pt](lr{0.125em}){1-1}\cmidrule[0.4pt](lr{0.125em}){2-2}\cmidrule[0.4pt](lr{0.125em}){3-3}\cmidrule[0.4pt](lr{0.125em}){4-4}\cmidrule[0.4pt](lr{0.125em}){5-5}
Model 1    & 49.11 $\pm$ 5.94e-5 & --- & --- & --- \\
\midrule
\multicolumn{1}{c}{} & \multicolumn{1}{c}{} & \multicolumn{1}{c}{$t_{\rm mem}$\, [s]} & \multicolumn{1}{c}{$t_{\rm mem}$\, [s]} & \multicolumn{1}{c}{$t_{\rm mem}$\, [s]} \\
& & $(\tau=\SI{0.5}{\second})$ & $(\tau=\SI{1.0}{\second})$ & $(\tau=\SI{2.5}{\second})$\\
\cmidrule[0.4pt](lr{0.125em}){3-3}\cmidrule[0.4pt](lr{0.125em}){4-4}\cmidrule[0.4pt](lr{0.125em}){5-5}
Model 2    & 14.53 $\pm$ 3.50e-5 & 6.29 $\pm$ 4.89e-2 & 6.50 $\pm$ 2.05e-2 & 6.20 $\pm$ 9.50e-1  \\
\midrule
\multicolumn{1}{c}{} & \multicolumn{1}{c}{} & \multicolumn{1}{c}{$t_{\rm mem}$\, [s]} & \multicolumn{1}{c}{$t_{\rm mem}$\, [s]} & \multicolumn{1}{c}{$t_{\rm mem}$\, [s]} \\
& & $(\tau=\SI{0.2}{\second})$ & $(\tau=\SI{0.4}{\second})$ & $(\tau=\SI{1.0}{\second})$\\
\cmidrule[0.4pt](lr{0.125em}){3-3}\cmidrule[0.4pt](lr{0.125em}){4-4}\cmidrule[0.4pt](lr{0.125em}){5-5}
Model 3    & 3.41 $\pm$ 2.78e-4 & 2.89 $\pm$ 3.37e-2 & 2.14 $\pm$ 3.64e-3 & 2.09 $\pm$ 3.45e-4  \\ 
\bottomrule
\label{Table1}
\end{tabular}
\end{table}

\begin{table}[h]
\caption{Correlation timescale $\tcorr$ and memory timescale $t_{\rm mem}$ for 
the experimental time-series of Hsp90 (here $q_0=0.25$ or $q_0=0.75$, and different $\tau$).
Values are obtained by fitting the long-time behaviour of $\mathcal{C}(t)$ and $\D(t)$, respectively, 
to the exponential $\simeq \exp(-t/t_i)$ where $i\in\{{\rm corr}, {\rm mem}\}$.}
\label{tab:exp}
\begin{tabular}{ccccc}
\toprule
\multicolumn{1}{c}{System} & \multicolumn{1}{c}{$t_{\rm corr}$\,[s]} & \multicolumn{1}{c}{$t_{\rm mem}$\, [s]} & \multicolumn{1}{c}{$t_{\rm mem}$\, [s]} & \multicolumn{1}{c}{$t_{\rm mem}$\, [s]} \\
& & $(\tau=\SI{1.0}{\second})$ & $(\tau=\SI{3.0}{\second})$ & $(\tau=\SI{5.0}{\second})$\\
\cmidrule[0.4pt](lr{0.125em}){1-1}\cmidrule[0.4pt](lr{0.125em}){2-2}\cmidrule[0.4pt](lr{0.125em}){3-3}\cmidrule[0.4pt](lr{0.125em}){4-4}\cmidrule[0.4pt](lr{0.125em}){5-5}
Hsp90 \\ ($q_0=0.25$)  & 90.00 $\pm$ 0.42 & 37.22 $\pm$ 14.69 & 45.98 $\pm$ 12.81 & 44.74 $\pm$ 17.20  \\
\cmidrule[0.4pt](lr{0.125em}){1-5}
\multicolumn{1}{c}{} & \multicolumn{1}{c}{} & \multicolumn{1}{c}{$t_{\rm mem}$\, [s]} & \multicolumn{1}{c}{$t_{\rm mem}$\, [s]} & \multicolumn{1}{c}{$t_{\rm mem}$\, [s]} \\
&  & $(\tau=\SI{1.0}{\second})$ & $(\tau=\SI{3.0}{\second})$ & $(\tau=\SI{5.0}{\second})$\\
\cmidrule[0.4pt](lr{0.125em}){3-3}\cmidrule[0.4pt](lr{0.125em}){4-4}\cmidrule[0.4pt](lr{0.125em}){5-5}
Hsp90 \\ ($q_0=0.75$) &  & 39.18 $\pm$ 16.71 & 47.92 $\pm$ 13.65 & 45.71 $\pm$ 18.46  \\
\bottomrule
\label{Table2}
\end{tabular}
\end{table}
\FloatBarrier
\newpage
\section*{Appendix: Memory analysis for simulated systems for $q_0=0.75$}
\FloatBarrier

\newpage
\section*{References}
\bibliography{MemoryPaper}

\end{document}